\documentclass{revtex4}

\usepackage{graphicx}
\newcommand{\nc}{\newcommand}
\def\nn{\nonumber\\}
\def\bea{\begin{eqnarray}}
\def\eea{\end{eqnarray}}
    
\nc{\braket}[1]{\langle\,{#1}\rangle}
\def\bra#1{\,\left\langle\, #1 \,\right\vert}
\def\ket#1{\,\left\vert\, #1 \,\right\rangle}

\def\pa{\partial}

  \def\R{\mathbf R}

  \newtheorem{Def}{Definition}[section]

  \newtheorem{Prob}[Def]{Problem}
  \newtheorem{Rule}[Def]{Rules}

\def\Cross{\unitlength.1em
  \begin{minipage}{15\unitlength}
    \begin{picture}(15,15)
      \put(15,15){\line(-1,-1){15}}
      \put(15,0){\line(-1,1){15}}
    \end{picture}
  \end{minipage}
}
\def\Smooth{\unitlength.1em
  \begin{minipage}{15\unitlength}
    \begin{picture}(15,15)
      \qbezier(0,0)(10,7.5)(0,15)
      \qbezier(15,0)(5,7.5)(15,15)
    \end{picture}
  \end{minipage}
}
\def\diaCrossP{\unitlength.1em
  \begin{minipage}{15\unitlength}
    \begin{picture}(15,15)
      \put(15,15){\vector(-1,-1){15}}
      \qbezier(15,0)(15,0)(10,5)
      \qbezier(5,10)(0,15)(0,15)
      \put(15,0){\vector(1,-1){0}}
    \end{picture}
  \end{minipage}
}
\def\diaCrossN{\unitlength.1em
  \begin{minipage}{15\unitlength}
    \begin{picture}(15,15)
      \put(0,15){\vector(1,-1){15}}
      \qbezier(0,0)(0,0)(5,5)
      \qbezier(10,10)(15,15)(15,15)
      \put(0,0){\vector(-1,-1){0}}
    \end{picture}
  \end{minipage}
}
\def\diaCrossPO{\unitlength.1em
  \begin{minipage}{15\unitlength}
    \begin{picture}(15,15)
      \put(0,0){\vector(1,1){15}}
      \qbezier(15,0)(15,0)(10,5)
      \qbezier(5,10)(0,15)(0,15)
      \put(0,15){\vector(-1,1){0}}
    \end{picture}
  \end{minipage}
}
\def\diaCrossNO{\unitlength.1em
  \begin{minipage}{15\unitlength}
    \begin{picture}(15,15)
      \put(15,0){\vector(-1,1){15}}
      \qbezier(0,0)(0,0)(5,5)
      \qbezier(10,10)(15,15)(15,15)
      \put(15,15){\vector(1,1){0}}
    \end{picture}
  \end{minipage}
}
\def\diaCrossPR{\unitlength.1em
  \begin{minipage}{15\unitlength}
    \begin{picture}(15,15)
      \put(0,15){\vector(1,-1){15}}
      \qbezier(0,0)(0,0)(5,5)
      \qbezier(10,10)(15,15)(15,15)
      \put(15,15){\vector(1,1){0}}
    \end{picture}
  \end{minipage}
}
\def\diaSmooth{\unitlength.1em
  \begin{minipage}{15\unitlength}
    \begin{picture}(15,15)
      \qbezier(0,0)(10,7.5)(0,15)
      \qbezier(15,0)(5,7.5)(15,15)
      \put(0,0){\vector(-1,-1){0}}
      \put(15,0){\vector(1,-1){0}}
    \end{picture}
  \end{minipage}
}
\def\diaL{\unitlength.1em
  \begin{minipage}{15\unitlength}
    \begin{picture}(15,15)
      \put(0,15){\vector(0,-1){15}}
    \end{picture}
  \end{minipage}
}
\def\diaR{\unitlength.1em
  \begin{minipage}{15\unitlength}
    \begin{picture}(15,15)
      \put(15,15){\vector(0,-1){15}}
    \end{picture}
  \end{minipage}
}
\def\diaCrossW{\unitlength.1em
  \begin{minipage}{14\unitlength}
    \begin{picture}(14,14)
      \put(7,7){\circle{8}}
      \qbezier(0,14)(0,14)(4,10)
      \qbezier(14,0)(14,0)(10,4)
      \qbezier(0,0)(0,0)(4,4)
      \qbezier(10,10)(14,14)(14,14)
    \end{picture}
  \end{minipage}
}
\def\diaCrossB{\unitlength.1em
  \begin{minipage}{14\unitlength}
    \begin{picture}(14,14)
      \put(7,7){\circle*{8}}
      \qbezier(0,14)(0,14)(5,9)
      \qbezier(14,0)(14,0)(9,5)
      \qbezier(0,0)(0,0)(5,5)
      \qbezier(9,9)(14,14)(14,14)
    \end{picture}
  \end{minipage}
}

\def\diaW{\unitlength.1em
  \begin{minipage}{10\unitlength}
    \begin{picture}(8,8)
      \put(4,4){\circle{8}}
    \end{picture}
  \end{minipage}
}
\def\diaB{\unitlength.1em
  \begin{minipage}{10\unitlength}
    \begin{picture}(8,8)
      \put(4,4){\circle*{8}}
    \end{picture}
  \end{minipage}
}

\def\diaC{\unitlength.1em
  \begin{minipage}{14\unitlength}
    \begin{picture}(14,14)
      \qbezier(0,0)(7,10)(14,0)
      \qbezier(0,14)(7,4)(14,14)
    \end{picture}
  \end{minipage}
}
\def\diaCap{\unitlength.1em
  \begin{minipage}{14\unitlength}
    \begin{picture}(14,14)
      \qbezier(0,0)(0,12)(7,12)
      \qbezier(7,12)(14,12)(14,0)
    \end{picture}
  \end{minipage}
}
\def\diaCup{\unitlength.1em
  \begin{minipage}{14\unitlength}
    \begin{picture}(14,14)
      \qbezier(0,14)(1,2)(7,2)
      \qbezier(7,2)(14,2)(14,14)
    \end{picture}
  \end{minipage}
}
\def\diaD{\unitlength.1em
  \begin{minipage}{14\unitlength}
    \begin{picture}(14,14)
      \qbezier(0,14)(0,2)(7,7)
      \qbezier(7,7)(14,12)(14,0)
    \end{picture}
  \end{minipage}
}

\begin{document}
\baselineskip=16pt
\title{Solving Infinite Kolam in Knot Theory}

\author{Yukitaka ISHIMOTO\footnote{e-mail: ishimoto@yukawa.kyoto-u.ac.jp}}
\address{Okayama Institute for Quantum Physics, 1-9-1 Kyouyama, Okayama 700-0015, Japan}
\date{Received 3 November 2006}

\begin{abstract}
\baselineskip=12pt
In southern India, there are traditional patterns of line-drawings encircling dots, called ``Kolam'', among which one-line drawings or the ``infinite Kolam'' provide very interesting questions in mathematics. 
For example, we address the following simple question: how many patterns of infinite Kolam can we draw for a given grid pattern of dots? 
The simplest way is to draw possible patterns of Kolam while judging if it is infinite Kolam. Such a search problem seems to be NP complete: almost all cases should be examined for a solution. However, it is certainly not. In this paper, we focus on diamond-shaped grid patterns of dots, (1-3-5-3-1) and (1-3-5-7-5-3-1) in particular. By using the knot-theory description of the infinite Kolam, we show how to find the solution, which inevitably gives a sketch of the proof for the statement ``infinite Kolam is not NP complete.'' Its further discussion will be given in the final section.
\end{abstract}

\begin{flushright}
OIQP-06-15
\end{flushright}

\maketitle

\noindent
Keywords: Kolam, knot theory, Morse link presentation, Temperley-Lieb algebra. 
\vspace{10mm}

\def\thesection{\arabic{section}}
\def\thesubsection{\arabic{section}.\arabic{subsection}}
\section{Introduction}

In southern India, there are many great female mathematicians who solve a complicated line pattern every morning, with white rice powder on the ground. The pattern is drawn around a grid pattern of dots so that the lines minimally encircle each dot, which is so called ``Kolam'' pattern in Tamil.  Among them, one-line drawings or the ``infinite Kolam'' introduced nicely in GERDES(1990), have some special meaning not only ethnologically but also mathematically (Fig.1). 
For example, we can address the following simple question: how many can we draw such patterns/diagrams for a given grid pattern of dots? 
For a particular grid pattern (1-3-1) in Fig.1., it is not difficult to find that there is only one infinite Kolam. But what happens when we have a bigger size of the grid pattern, say $3 \times 3$ or $10 \times 10$?
\begin{figure}[h]
\centering
\includegraphics[width=11.0cm, height=32mm]{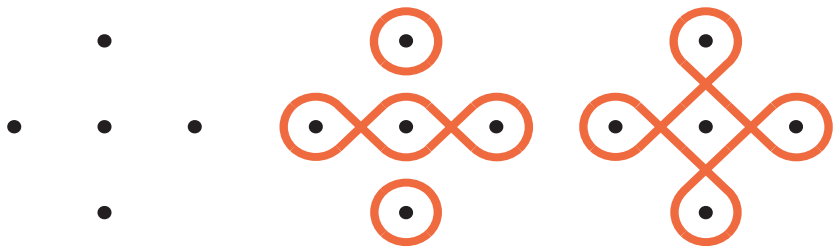}
\caption{Left is a grid pattern (1-3-1) of dots. Center is an example of Kolam, while Right is the only example of infinite Kolam for this (1-3-1) pattern.}
\end{figure}

As one can easily confirm by hand, the problem turns out to be very hard, if one simply follows the general rules for Kolam, which will be given in the next section, and judge if thus drawn Kolam is infinite Kolam or not.
One may think that such a search problem is even NP complete: substantial computational effort is required for finding the solution. However, there seem to be some ways or algorithms to reduce such effort.

One way is to use the mirror curve method. 
Since more than a decade ago, it has been shown that the infinite Kolam can be constructed as mirror curves in regular lattices (see for example GERDES(1990),JABLAN(to appear) and references therein). In fact, we can relatively easily construct some of the simplest examples of infinite Kolam by this method. This might lead us to the simplest way to construct all the possible patterns of the infinite Kolam, or to the statement that ``infinite Kolam is not NP complete.'' However, it seems to be still very difficult to show them by this method. The answer to our question has not been shown yet by this method, nor has a proof for the statement.

There is another way to reduce the task, which is a knot theoretical way. When we remove the dots from the Kolam patterns, the remaining line-drawings are merely the two-dimensional projections of corresponding link diagrams in knot theory (KAUFFMAN,1991). Therefore, the Kolam patterns can be described and analyzed by the techniques developed in or based on knot theory. The link diagram is such a diagram that consists of one or more knots --- called components of the link. Hence, the number of one component link diagrams gives the answer to our question.

In this paper, we focus on the so-called ``diamond Kolam'' and show the way towards the answer. The diamond Kolam is such Kolam whose grid patterns of dots are all diamond-shaped. The simplest examples of diamond Kolam are shown in Fig.1. If we successively write the number of dots in each row, those grid patterns can be represented by a sequence of numbers from the top row to the bottom as follows:
\begin{center}
(1-3-1), (1-3-5-3-1), (1-3-5-7-5-3-1), and so on. 
\end{center}
We use this notation in what follows. 
By using the knot-theory description of the Kolam, we show how to find the solution, $i.e.$, the infinite Kolam, and give the answer to our question in the diamond Kolam case: 1 for (1-3-1), 240 for (1-3-5-3-1), and 11,661,312 for (1-3-5-7-5-3-1) without excluding symmetric solutions. The first two entries match the results in (YANAGISAWA,2006), and the final one has just been confirmed in another way by Yanagisawa, where he examined all the cases of (1-3-5-7-5-3-1) Kolam.
Unfortunately, we have not reach the general solution of the problem, that is, the general expression of the numbers for any size of the diamond Kolam. Not to mention for any grid pattern of dots. However, we have shown the key steps for the general solution, 
which inevitably gives a sketch of the proof for the statement ``infinite Kolam is not NP complete.'' A further discussion will be given in the final section from this mathematical point of view.
Note that the procedure given here is not restricted to the diamond Kolam, and therefore it can be applied to any other shape or problem in the universe of Kolam.

\section{Problem and the method}

In this section, we first present the problem together with the general rules for Kolam patterns. Secondly, the problem defined on the grid pattern of dots is translated into the problem on the corresponding checker-pattern lattice. Then, the Kolam patterns are replaced by mathematical expressions with the Temperley-Lieb algebra with some modification. During this process, we show how to express the Kolam patterns in the Morse link presentations of knot theory. Finally, we introduce the upper-lower diagrammatic decomposition of the Kolam, and its mathematical treatment in linear algebra, {\it i.e.}, vectors, matrices, and metric. The examples for (1-3-1), (1-3-5-3-1) are also given in each subsection. The precise results for (1-3-1), (1-3-5-3-1), (1-3-5-7-5-3-1) are given in the next section.

\subsection{Problem, Rules, and Translation}

First, we reintroduce the problem: \\[0pt]

\fbox{
\begin{minipage}{13cm}
\begin{Prob}
(the infinite Kolam counting)\\
For a given grid pattern of dots on a square lattice, find the number of one-line drawings which satisfy the general rules for Kolam.
\end{Prob}
\end{minipage}
}\\[5pt]

There are a variety of extensions or modifications to a minimal set of the rules for Kolam (GERDES,1990;NAGATA et al.,2004;NAGATA et al.,2006). Therefore, to be more precise, we define here the general rules for Kolam, which are the rules for the lines in Kolam. Note that the dots are placed on sites of a square lattice and `edge' is defined by a straight line segment connecting two nearest neighbourhood sites. 
\\[0pt]

\fbox{
\begin{minipage}{13cm}
\begin{Rule}
(the general rules for Kolam)
\begin{enumerate}
\item
The lines are smooth and closed, and their crossings are allowed.
\item
The lines pass through middle points of the edges of the lattice, and only through such points nearest to the dots. At most, two lines pass at each point. 
\item
Each dot must be isolated by, at most, four lines passing through only middle points of the edges nearest to the dot. 
\end{enumerate}
\end{Rule}
\end{minipage}
}\\[15pt]
The closed line simply means no ends.
The first rule mentions the crossing, but its order is generally not questionable because the patterns are usually drawn by fine granular objects. 
The second rule prescribes the basic paths for the line-drawings, which also implies that the paths are smooth short-cuts connecting the middle points nearest to each other. The rule further means that any other point of the edges are not allowed to be passed.
In practice, the uncrossed two neighbouring lines are not exactly passing through the middle points of the edges but rather through their neighbourhood points. Nonetheless, we define them as above to simplify the rules.
The third rule restricts or attracts the paths just around the dots. With the rules, we have basically six patterns of lines around a dot (Fig.2).
\begin{figure}[h]
\centering
\includegraphics[width=11.0cm, height=2.0cm]{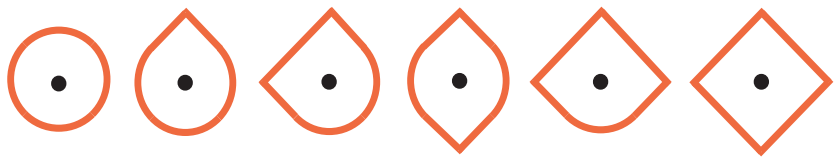}
\caption{The basic six patterns of lines around each dot, excluding symmetric patterns by mirror symmetries and rotations.}
\end{figure}
The one-line drawing should also be defined here: a single closed smooth line with some self-crossings. It is intrinsically different from but historically influenced by the definition of Euleric curves.
Note that, being inspired by NAGATA et al.(2004), NAGATA et al.(2006), and YANAGISAWA(2006), we further focus on the diamond shaped grid pattern of dots, the diamond Kolam which was mentioned in the previous section. 

Now, we are ready to translate Problem 2.1 defined on a grid pattern of dots into a problem on a checker-pattern lattice.
According to Rules 2.2, the vertex positions of the drawing (including crossings) are restricted to the neighbourhoods of the middle points of the edges.
The rules further separate the possible vertices into two types: one is just below the dots (we call this `black site') and the other is just aside the dots (`white site'). Then, one obtains the alternating B/W checker pattern lattice, somehow dual to the original (Fig.3). 
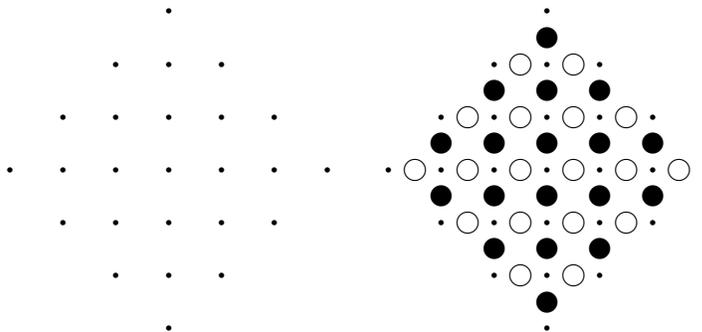
\begin{figure}[h]
\begin{center}
\begin{picture}(140,140)
   \put(70,10){\circle*{2}}   \put(50,30){\circle*{2}}
   \put(70,30){\circle*{2}}   \put(90,30){\circle*{2}}
   \put(30,50){\circle*{2}}   \put(50,50){\circle*{2}}
   \put(70,50){\circle*{2}}   \put(90,50){\circle*{2}}
   \put(110,50){\circle*{2}}   \put(10,70){\circle*{2}}
   \put(30,70){\circle*{2}}   \put(50,70){\circle*{2}}
   \put(70,70){\circle*{2}}   \put(90,70){\circle*{2}}
   \put(110,70){\circle*{2}}   \put(130,70){\circle*{2}}
   \put(30,90){\circle*{2}}   \put(50,90){\circle*{2}}
   \put(70,90){\circle*{2}}   \put(90,90){\circle*{2}}
   \put(110,90){\circle*{2}}   \put(50,110){\circle*{2}}
   \put(70,110){\circle*{2}}   \put(90,110){\circle*{2}}
   \put(70,130){\circle*{2}}
\end{picture}
\begin{picture}(140,140)
   \put(70,10){\circle*{2}}   \put(50,30){\circle*{2}}
   \put(70,30){\circle*{2}}   \put(90,30){\circle*{2}}
   \put(30,50){\circle*{2}}   \put(50,50){\circle*{2}}
   \put(70,50){\circle*{2}}   \put(90,50){\circle*{2}}
   \put(110,50){\circle*{2}}   \put(10,70){\circle*{2}}
   \put(30,70){\circle*{2}}   \put(50,70){\circle*{2}}
   \put(70,70){\circle*{2}}   \put(90,70){\circle*{2}}
   \put(110,70){\circle*{2}}   \put(130,70){\circle*{2}}
   \put(30,90){\circle*{2}}   \put(50,90){\circle*{2}}
   \put(70,90){\circle*{2}}   \put(90,90){\circle*{2}}
   \put(110,90){\circle*{2}}   \put(50,110){\circle*{2}}
   \put(70,110){\circle*{2}}   \put(90,110){\circle*{2}}
   \put(70,130){\circle*{2}}
   \put(70,20){\circle*{8}}   \put(50,40){\circle*{8}}
   \put(70,40){\circle*{8}}   \put(90,40){\circle*{8}}
   \put(30,60){\circle*{8}}   \put(50,60){\circle*{8}}
   \put(70,60){\circle*{8}}   \put(90,60){\circle*{8}}
   \put(110,60){\circle*{8}}   \put(30,80){\circle*{8}}
   \put(50,80){\circle*{8}}   \put(70,80){\circle*{8}}
   \put(90,80){\circle*{8}}   \put(110,80){\circle*{8}}
   \put(50,100){\circle*{8}}   \put(70,100){\circle*{8}}
   \put(90,100){\circle*{8}}   \put(70,120){\circle*{8}}
   \put(60,30){\circle{8}}   \put(80,30){\circle{8}}
   \put(40,50){\circle{8}}   \put(60,50){\circle{8}}
   \put(80,50){\circle{8}}   \put(100,50){\circle{8}}
   \put(20,70){\circle{8}}   \put(40,70){\circle{8}}
   \put(60,70){\circle{8}}   \put(80,70){\circle{8}}
   \put(100,70){\circle{8}}   \put(120,70){\circle{8}}
   \put(40,90){\circle{8}}   \put(60,90){\circle{8}}
   \put(80,90){\circle{8}}   \put(100,90){\circle{8}}
   \put(60,110){\circle{8}}   \put(80,110){\circle{8}}
\end{picture}
\caption{The (1-3-5-7-5-3-1) dot pattern of the diamond Kolam (Left) and the superposition of the dot pattern and the B/W checker pattern lattice (Right).}
\end{center}
\end{figure}
The black site takes either `cross' or `cup and cap', while the white takes either `cross' or `recoil':
\bea
{\rm Black~ site:~} 
\diaCrossB = \Cross {\rm ~or~} \diaC \;, \nn
{\rm White~ site:~} 
\diaCrossW = \Cross {\rm ~or~} \Smooth \;.
\eea
Accordingly, the total number of Kolam patterns for a given grid pattern of dots is given by
\bea
  2^{D} , 
\eea
where $D$ denotes the number of the black and white sites:
\bea
 D = \left\{ \begin{array}{ll}
       4 & {\rm for~ \mbox{(1-3-1)}} \\
      16 & {\rm for~ \mbox{(1-3-5-3-1)}} \\
      36 & {\rm for~ \mbox{(1-3-5-7-5-3-1)}} \\
      4 N^2 & {\rm for~ \mbox{(1-3- $\cdots$ -$(2N+1)$- $\cdots$ -3-1)}}, N=1,2,3,\cdots.
      \end{array} \right. .
\eea
This obviously shows that the number of cases increases astronomically as the size of the diamond Kolam grows.

Following Rules 2.2, there is left no choice for the lines except at the black and white sites. Namely, the lines except at the B/W sites are the ones connecting the neighbourhood B/W sites and the ones encircling boundary dots. Consequently, the dots are found to be irrelevant after this translation so that they can be removed from the patterns, leaving only diagrams of the lines.
Hence, the problem can be reduced to that of the knot group acting on the given vertical lines (14 lines for the above case) with caps and cups on the boundary dots (Fig.4). In the next subsection, we describe the diagrams and analyze them in knot theory. Note that this translation can be shown equivalent to the mirror curves (GERDES,1990;JABLAN, to appear) for this particular grid pattern.

\subsection{A knot-theoretical description and a modification to the Temperley-Lieb algebra}

The diagrams in Kolam patterns are now extracted in the previous subsection as the checker pattern lattice with the fixed lines. This family of diagrams can be represented by braid representatives with cups and caps or Morse link presentations in knot theory (Fig.4).
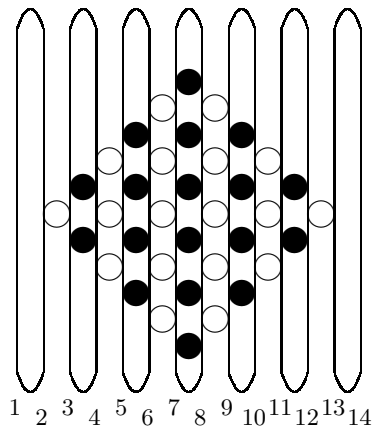
\begin{figure}[h]
\begin{center}
\begin{picture}(140,150)
   \put(5,10){\line(0,1){130}}
   \put(15,10){\line(0,1){130}}
   \put(25,10){\line(0,1){130}}
   \put(35,10){\line(0,1){130}}
   \put(45,10){\line(0,1){130}}
   \put(55,10){\line(0,1){130}}
   \put(65,10){\line(0,1){130}}
   \put(75,10){\line(0,1){130}}
   \put(85,10){\line(0,1){130}}
   \put(95,10){\line(0,1){130}}
   \put(105,10){\line(0,1){130}}
   \put(115,10){\line(0,1){130}}
   \put(125,10){\line(0,1){130}}
   \put(135,10){\line(0,1){130}}
     \qbezier(5,140)(10,155)(15,140)
     \qbezier(25,140)(30,155)(35,140)
     \qbezier(45,140)(50,155)(55,140)
     \qbezier(65,140)(70,155)(75,140)
     \qbezier(85,140)(90,155)(95,140)
     \qbezier(105,140)(110,155)(115,140)
     \qbezier(125,140)(130,155)(135,140)
     \qbezier(5,10)(10,-5)(15,10)
     \qbezier(25,10)(30,-5)(35,10)
     \qbezier(45,10)(50,-5)(55,10)
     \qbezier(65,10)(70,-5)(75,10)
     \qbezier(85,10)(90,-5)(95,10)
     \qbezier(105,10)(110,-5)(115,10)
     \qbezier(125,10)(130,-5)(135,10)
   \put(70,20){\circle*{10}}   \put(50,40){\circle*{10}}
   \put(70,40){\circle*{10}}   \put(90,40){\circle*{10}}
   \put(30,60){\circle*{10}}   \put(50,60){\circle*{10}}
   \put(70,60){\circle*{10}}   \put(90,60){\circle*{10}}
   \put(110,60){\circle*{10}}   \put(30,80){\circle*{10}}
   \put(50,80){\circle*{10}}   \put(70,80){\circle*{10}}
   \put(90,80){\circle*{10}}   \put(110,80){\circle*{10}}
   \put(50,100){\circle*{10}}   \put(70,100){\circle*{10}}
   \put(90,100){\circle*{10}}   \put(70,120){\circle*{10}}
   \put(60,30){\circle{10}}   \put(80,30){\circle{10}}
   \put(40,50){\circle{10}}   \put(60,50){\circle{10}}
   \put(80,50){\circle{10}}   \put(100,50){\circle{10}}
   \put(20,70){\circle{10}}   \put(40,70){\circle{10}}
   \put(60,70){\circle{10}}   \put(80,70){\circle{10}}
   \put(100,70){\circle{10}}   \put(120,70){\circle{10}}
   \put(40,90){\circle{10}}   \put(60,90){\circle{10}}
   \put(80,90){\circle{10}}   \put(100,90){\circle{10}}
   \put(60,110){\circle{10}}   \put(80,110){\circle{10}}
   \put(2,-6){1}
   \put(12,-10){2}
   \put(22,-6){3}
   \put(32,-10){4}
   \put(42,-6){5}
   \put(52,-10){6}
   \put(62,-6){7}
   \put(72,-10){8}
   \put(82,-6){9}
   \put(90,-10){10}
   \put(100,-6){11}
   \put(110,-10){12}
   \put(120,-6){13}
   \put(130,-10){14}
\end{picture}
\end{center}
\caption{Braid representatives with cups and caps inserted, or Morse link presentations of the family of the Kolam diagrams. For convenience, numbers at the bottom are assigned to the vertical lines.}
\end{figure}
That is to say, the Kolam diagrams of (1-3-...-($2N+1$)-...-3-1) are intrinsically the $4(2N+1)$ vertical strings acted by either \Cross or \diaC. 
Therefore, it is natural to write these diagrams by the algebra known as the Temperley-Lieb algebra with the braid group (KAUFFMAN,1991;TEMPERLEY et al.,1971), which contains the permutations \Cross, the cap \diaCap, and the cup \diaCup. The essence of the algebra is just the usual topologically equivalent operations on knots, so it can be easily expressed by diagrams. For example, the famous Reidemeister moves are realised in the algebra, and are expressed as follows:\\

\underline{\bf Reidemeister moves}
\begin{enumerate}
 \def\theenumi{\roman{enumi}}
 \def\labelenumi{(\theenumi)}
\item 
$ \begin{array}{c} 
     \phantom{\diaCrossP} \diaCap \phantom{\diaSmooth} \\
       {\diaCrossP} \phantom{\diaCap} {\Smooth} \\
     \phantom{\diaCrossP} \diaCup \phantom{\diaSmooth} 
  \end{array}
  = \Smooth $ ,
$ \begin{array}{c} 
     \phantom{\diaCrossP} \diaC \phantom{\diaSmooth} \\
     \phantom{\diaCrossP} \diaCrossPR \phantom{\diaSmooth} 
  \end{array}
  = \diaC $ ,
\item 
$\begin{array}{c} \diaCrossP \\ \diaCrossN \end{array} = \diaSmooth 
\;= \begin{array}{c} \diaCrossN \\ \diaCrossP \end{array} $ , 
$\begin{array}{c} 
      \phantom{\diaCrossP} \diaCap \phantom{\diaCrossN} \\
      \diaCrossP \phantom{\diaC} \diaCrossNO \\
      \phantom{\diaCrossP} \diaCup \phantom{\diaCrossN}
 \end{array}
 = \diaC = 
\begin{array}{c} 
      \phantom{\diaCrossP} \diaCap \phantom{\diaCrossN} \\
      \diaCrossN \phantom{\diaC} \diaCrossPO \\
      \phantom{\diaCrossP} \diaCup \phantom{\diaCrossN}
 \end{array}
$ .
\item 
$\begin{array}{c} 
      \diaCrossN \phantom{\diaCrossPO} \\
      \diaL \diaCrossN \\
      \diaCrossP \phantom{\diaCrossPO}
 \end{array}
 =
 \begin{array}{c} 
      \phantom{\diaL} \diaCrossP \\
      \diaCrossN \diaR \\
      \phantom{\diaL} \diaCrossN 
 \end{array}
$ .\\
\end{enumerate}
Once the diagrams are written down by the algebra, they can naturally be reduced to topologically equivalent simple diagrams and their corresponding algebraic polynomials in an appropriate way. However, the known polynomials, such as the Alexander-Conway polynomial or the Jones polynomial, are only powerful for computing the knot invariants but not for counting the number of one-line drawings. Therefore, we need another trick to extract the number from such polynomials of links. 

Here, we introduce a modification to the algebra so that it gives a polynomial in link components rather than for the knot. The modification consists of the full decomposition of $\;\diaC$ operator in the Temperley-Lieb algebra into caps and cups, and the projection that eliminates the signs of crossings in diagrams.
The diagrammatic features of this projection are drawn below:
\bea
\begin{array}{c} \diaC \\ \Cross \end{array} &=& \diaC \;, \quad 
\Cross \Smooth \;=\; \Smooth , 
\nn
\begin{array}{c} \Cross \\ \Cross \end{array} &=& \Smooth \;, \quad 
\Cross \Cross \;=\; \diaC , 
\nn
\begin{array}{c} \diaCap \\ \Cross \end{array} &=& \diaCap \,, \quad
\begin{array}{c} \diaC \\ \Cross \end{array} = \diaC \,, \quad
\begin{array}{c} \Cross \\ \diaCup \end{array} = \diaCup \,,\quad
\begin{array}{c} \Cross \\ \diaC \end{array} = \diaC \,. \quad
\eea
These operations obviously preserve the number of link components.
Before deriving the polynomial, let us first fix the mathematical notations for this Morse link presentation.

Define the vertical strings by numbering them from left to right starting from one to $2(2N+1)$ as in Fig.4.
Then, define the operator $\Cross_{\hspace{-12pt}i~~~j}$ which represents the permutation between the $i$-th and $j$-th strings.
Similarly, we define the operators $\diaCap_{\hspace{-12pt}i~~~j}$\,,\, $\diaCup^{\hspace{-12pt}i~~~j}$, and also $\diaD^{\hspace{-12pt}i}_{~j}$ which connects the $i$-th string from above to the $j$-th string downwards. By these definitions, the black site is given by $\Cross_{\hspace{-12pt}i~~~i+1}$ ~or~ $\diaCup^{\hspace{-12pt}i~~~i+1} \diaCap_{\hspace{-12pt}i~~~i+1}$, while the white one is $\Cross_{\hspace{-12pt}i~~~i+1}$ ~or~ $\Smooth_{\hspace{-12pt}i~~~i+1}=\diaD^{\hspace{-12pt}i}_{~i} \diaD^{\hspace{-12pt}i+1}_{~i+1}$. The \Smooth operation is simply an identity operation to this algebra so that we write this by ${\mathbf 1}$.
In order to view this algebra in a more convenient form, we change the diagrammatic order of top-to-bottom to the algebraic order of left-to-right. For example,
\bea
 \begin{array}{c} {\Cross}\\ {\Cross} \\ i~~j \end{array}
  = \Cross_{\hspace{-12pt}i~~~j} \; \Cross_{\hspace{-12pt}i~~~j }  , \quad 
 \begin{array}{c} \diaC \\ {\phantom{n}}^i~~{}^{i+1} \end{array}
  = \diaCup^{\hspace{-12pt}i~~~i+1} \diaCap_{\hspace{-12pt}i~~~i+1}
\eea
Finally, we add the contraction between the cap and cup as follows.
\bea
    \diaCap_{\hspace{-12pt}i~~~j} \diaCup^{\hspace{-12pt}i~~~j} = a, 
\eea
where $a$ can take any number so far.
Mathematically, there exist already the corresponding set of symbols:
\bea
\Cross_{\hspace{-12pt}i~~~j} &=& \sigma_{i,j} , \quad
\diaCup^{\hspace{-12pt}i~~~j} \,=\, \cup_{i,j} , \nn
\diaD^{\hspace{-12pt}i}_{~j} &=& \delta^i_j , \quad\;\;
\diaCap_{\hspace{-12pt}i~~~j} \,=\, \cap_{i,j} .
\eea
By introducing another symbol $U_{i,j}$ such that $U_{i,j} \equiv \diaCup^{\hspace{-12pt}i~~~j} \diaCap_{\hspace{-12pt}i~~~j}$, the complete set of the modified Temperley-Lieb algebra is given in the algebraic form in the appendix A.

In knot theory, it is known that any knot or link can be represented by the Morse link presentation and by a sequence of operators --- a monomial in the relevant algebra. Indeed, one can see that any Kolam diagram can be represented by a monomial of the four operators with the identity ${\mathbf 1}$ in our case, and s/he can derive a number as a result, following the above algebraic rules. For example, the infinite Kolam in Fig.1 gives the following monomial: 
\bea&&
\braket{
  \diaCap_{\hspace{-12pt}1~~~2} \; 
  \diaCap_{\hspace{-12pt}3~~~4} \; 
  \diaCap_{\hspace{-12pt}5~~~6} \; 
  \Cross_{\hspace{-12pt}3~~~4} \; 
  \Cross_{\hspace{-12pt}2~~~3} \; 
  \Cross_{\hspace{-12pt}4~~~5} \; 
  \Cross_{\hspace{-12pt}3~~~4} \; 
  \diaCup^{\hspace{-12pt}1~~~2} \; 
  \diaCup^{\hspace{-12pt}3~~~4} \; 
  \diaCup^{\hspace{-12pt}5~~~6} 
}
\nn&&
  = 
\braket{
  \diaCap_{\hspace{-12pt}1~~~2} \; 
  \diaCap_{\hspace{-12pt}3~~~4} \; 
  \Cross_{\hspace{-12pt}3~~~4} \; 
  \diaCap_{\hspace{-12pt}5~~~6} \; 
  \Cross_{\hspace{-12pt}2~~~3} \; 
  \Cross_{\hspace{-12pt}4~~~5} \; 
  \diaCup^{\hspace{-12pt}1~~~2} \; 
  \Cross_{\hspace{-12pt}3~~~4} \; 
  \diaCup^{\hspace{-12pt}3~~~4} \; 
  \diaCup^{\hspace{-12pt}5~~~6} 
}
\nn&&
  = 
  \braket{ \diaCap_{\hspace{-12pt}1~~~2} \; 
  \diaCap_{\hspace{-12pt}3~~~4} \; 
  \diaCap_{\hspace{-12pt}5~~~6} \; 
  \Cross_{\hspace{-12pt}2~~~3} \; 
  \Cross_{\hspace{-12pt}4~~~5} \; 
  \diaCup^{\hspace{-12pt}1~~~2} \; 
  \diaCup^{\hspace{-12pt}3~~~4} \; 
  \diaCup^{\hspace{-12pt}5~~~6} 
  }
\nn&&
  = 
  \braket{ \diaCap_{\hspace{-12pt}3~~~4} \; 
  \diaCup^{\hspace{-12pt}3~~~4} \; 
  }
\nn&&
  = a .
\eea
The bracket $\braket{...}$ is to denote the algebraic expression of diagrams, which will later be linked to the notion of bra and ket states of Dirac.
From the first line to the second, the monomial is re-ordered for computation.
From the second to the third, the third and fourth lines of Eq.(4) are used. From the third to the fourth, Reidemeister move (i) is used twice.
Similarly, the non-infinite Kolam in Fig.1 reads
\bea&&
\braket{
  \diaCap_{\hspace{-12pt}1~~~2} \; 
  \diaCap_{\hspace{-12pt}3~~~4} \; 
  \diaCap_{\hspace{-12pt}5~~~6} \; 
  \diaC_{\hspace{-12pt}3~~~4} \; 
  \Cross_{\hspace{-12pt}2~~~3} \; 
  \Cross_{\hspace{-12pt}4~~~5} \; 
  \diaC_{\hspace{-12pt}3~~~4} \; 
  \diaCup^{\hspace{-12pt}1~~~2} \; 
  \diaCup^{\hspace{-12pt}3~~~4} \; 
  \diaCup^{\hspace{-12pt}5~~~6} 
}
\nn&&
  = 
  \braket{ \diaCap_{\hspace{-12pt}3~~~4} \; 
  \diaC_{\hspace{-12pt}3~~~4} \; 
  \diaC_{\hspace{-12pt}3~~~4} \; 
  \diaCup^{\hspace{-12pt}3~~~4} \; 
  }
\nn&&
  = 
  \braket{ \diaCap_{\hspace{-12pt}3~~~4} \; 
  \diaCup^{\hspace{-12pt}3~~~4} \; \diaCap_{\hspace{-12pt}3~~~4} \; 
  \diaCup^{\hspace{-12pt}3~~~4} \; \diaCap_{\hspace{-12pt}3~~~4} \; 
  \diaCup^{\hspace{-12pt}3~~~4} \; 
  }
\nn&&
  = a^3 .
\eea
From the first line to the second, Reidemeister move (i) is used twice.
From the second to the third, the operator decomposition by caps and cups are applied.
Hence, the former case -- infinite Kolam contains only one component, while the latter contains three components.

From this lesson, it is now trivial that the diagram is one-line drawing if the corresponding monomial is reduced to the number: $a$. Besides, it is clear that if the diagram contains $n$ loops, or $n$ components, then the corresponding monomial becomes $a^n$. If we calculate all the possible diagrams at once, it appears to be a polynomial in $a$. In such a situation, the number of one-component diagrams is simply given by the coefficient of the first order term of $a^1$ in the polynomial. There are various ways of obtaining this. For example, obtain the polynomial and differentiate it in a, then substitute $a=0$ into it. Or, define $a$ is grassmannian, so that $a\ne 0$ and $a^2=0$, so the diagram of more than one component vanish from the polynomial. It should be noted here that the latter implies a further reduction of the computational effort which we will mention later. 
At any rate, we should first clarify how to obtain the polynomial.

For the derivation of the polynomial, we introduce the following two operators at the black and white sites:
\bea
  \diaB_{i} &=& \Cross_{\hspace{-12pt}i~~~i+1} + \diaCup^{\hspace{-12pt}i~~~i+1} \diaCap_{\hspace{-12pt}i~~~i+1}, \nn
  \diaW_{i} &=& \Cross_{\hspace{-12pt}i~~~i+1} + {\mathbf 1} .
\eea
In addition, we introduce the top state (TOP state) and the bottom state (BOT state), using the bra and ket notion of Dirac:
\bea
  &{\rm Top~ state:~}& \bra{TOP_{2N+1}} \equiv 
  \bra{\diaCap_{\hspace{-12pt}1~~~2} \cdots ~~~~~~~~ \diaCap_{\hspace{-32pt} {}^{4N+1} ~~~~{}^{4N+2}} }
\nn
  &{\rm Bottom~ state:~}& \ket{BOT_{2N+1}} \equiv 
  \ket{\diaCup^{\hspace{-12pt}1~~~2} \cdots  ~~~~~~~~ \diaCup^{\hspace{-32pt} {}^{4N+1} ~~~~{}^{4N+2}} } \;,
\eea
for the (1-3-...-(2N+1)-...-3-1) diamond Kolam. Notice that they are trivially given by the maximum number of dots in rows. Say, for the (3-3-3) Kolam, they are $\bra{TOP_{3}}$ and $\ket{BOT_3}$.
Consequently, the polynomial $P$ of the diamond Kolam is given by:
\bea
  P\left[ \mbox{(1-3-...-(2N+1)-...-3-1)} \right]
  \equiv
  \bra{TOP_{2N+1}} \diaB \diaW^2 \diaB^3 \cdots \diaW^{2N} \cdots \diaB^3 \diaW^2 \diaB \ket{BOT_{2N+1}} , 
\eea
where the lower suffixes of the black and white circle operators are neglected for brevity.
This reveals another interpretation of the diagram: the black and white circle operators act on the top state and transform it to another state, then the inner product between the transformed state and the bottom state produce a number. We will see this in the rest of this section.

\subsection{Upper-Lower decomposition and state space}

In this subsection, we examine the upper half of the diagram in an algebraic and knot theoretical way and show that they form a linear space of states.

As it may be noticed, the family of diagrams expressed in Fig.4 and by Eq.(11) are totally symmetric. The symmetries include up-down symmetry, left-right symmetry: seven symmetries in total. 
If one focuses on the up-down symmetry, one can find that the calculation of the upper half diagram is just the same as of the lower half. Indeed, we can see this more rigorously by defining the transposition of the states and the operators:
\bea&&
(O_1 O_2)^T = O_2^T O_1^T, \quad \left(O_1^T \right)^T = O_1 , \nn &&
  \Cross^T = \Cross, \quad  
  \diaCap^T = \diaCup, \quad
  \diaCup^T = \diaCap, \nn&&
  U^T = U,\quad  {\mathbf 1}^T = {\mathbf 1} , 
\eea
where $O_i$ stands for any operator of the algebra and the suffixes are neglected as the transposition is defined under which they do not change.
The definition means that
\bea&&
  \bra{TOP_N}^T = \ket{BOT_N}, \quad
  \{\diaB \diaW^2 ... \diaB^{n-1}\}^{T} = \{\diaB^{n-1} ... \diaW^2 \diaB\} , \quad 
\nn&&
  {\rm T}: \bra{TOP_N} \diaB \diaW^2 ... \diaB^{N-2}  
   \longleftrightarrow 
           \diaB^{N-2} ... \diaW^2 \diaB \ket{BOT_N} .
\eea
Hence, the problem is now further reduced to finding the form of the upper half diagram and calculating its inner product with its transposition acted by $\diaW^{N-1}$ in the middle of the diagram: 
\bea
  P\left[ \mbox{(1-3-...-N-...-3-1)} \right]
  &=& \bra{{\rm Upper}_N} \diaW^{N-1} \ket{{\rm Lower}_N} ,
\eea
where
\bea&&
  \bra{{\rm Upper}_N} \equiv \bra{TOP_N} \diaB \diaW^2 ... \diaB^{N-2} ,
  \nn&&
  \ket{{\rm Lower}_N} \equiv \bra{{\rm Upper}_N}^T . 
\eea
Diagrammatically, for instance, $\bra{{\rm Upper}_7}$ is given by
\bea
\bra{{\rm Upper}_7} = 
\begin{minipage}{6cm}
\begin{picture}(140,80)
   \put(5,10){\line(0,1){60}}   \put(15,10){\line(0,1){60}}
   \put(25,10){\line(0,1){60}}   \put(35,10){\line(0,1){60}}
   \put(45,10){\line(0,1){60}}   \put(55,10){\line(0,1){60}}
   \put(65,10){\line(0,1){60}}   \put(75,10){\line(0,1){60}}
   \put(85,10){\line(0,1){60}}   \put(95,10){\line(0,1){60}}
   \put(105,10){\line(0,1){60}}   \put(115,10){\line(0,1){60}}
   \put(125,10){\line(0,1){60}}   \put(135,10){\line(0,1){60}}
     \qbezier(5,70)(10,85)(15,70)     \qbezier(25,70)(30,85)(35,70)
     \qbezier(45,70)(50,85)(55,70)     \qbezier(65,70)(70,85)(75,70)
     \qbezier(85,70)(90,85)(95,70)     \qbezier(105,70)(110,85)(115,70)
     \qbezier(125,70)(130,85)(135,70)
   \put(70,60){\circle*{10}}   \put(50,40){\circle*{10}}
   \put(70,40){\circle*{10}}   \put(90,40){\circle*{10}}
   \put(30,20){\circle*{10}}   \put(50,20){\circle*{10}}
   \put(70,20){\circle*{10}}   \put(90,20){\circle*{10}}
   \put(110,20){\circle*{10}}  
   \put(60,50){\circle{10}}   \put(80,50){\circle{10}}
   \put(40,30){\circle{10}}   \put(60,30){\circle{10}}
   \put(80,30){\circle{10}}   \put(100,30){\circle{10}}
   \put(2,-6){1}   \put(12,-10){2}
   \put(22,-6){3}   \put(32,-10){4}
   \put(42,-6){5}   \put(52,-10){6}
   \put(62,-6){7}   \put(72,-10){8}
   \put(82,-6){9}   \put(90,-10){10}
   \put(100,-6){11}   \put(110,-10){12}
   \put(120,-6){13}   \put(130,-10){14}
\end{picture}
\end{minipage}
.
\\\nonumber
\eea

Each diagram in $\bra{{\rm Upper}_N}$ has $2N$ legs, and it is reduced to the simplest topologically equivalent diagram, that is, $\bra{\cap_{i_1,i_2} \cdots \cap_{i_{N-1}, i_{N}}}$ of $N$ pairs for $i_j \ne i_{k\ne j}$. We call this an ``irreducible diagram.'' Such irreducible diagrams are distinguished from other topologically inequivalent irreducible diagrams, and therefore they form a linear space. We call the irreducible diagram the ``basis state'' and the linear space is called the ``state space.'' The top state $\bra{TOP_N}$ is one of the basis states and is a base of this state space. The dimension of the state space is given by the number of combinations of $N$ pairs of $2N$ legs: $2^N\, \Gamma(N+1/2)/\Gamma(1/2)$. The upper half diagram is now translated to a state in this state space, which is uniquely given by a linear combination of the basis states. The dual of the state space is apparently of the lower half.
An explicit example will be given in the following sections.
Note that the coefficient of a basis state gives the occurrence of the basis state in the upper half diagram, which is usually given by a polynomial in $a$. If we set $a=0$ in the upper half diagram, it cuts down the diagrams with trivial knots and leaves only the candidates for the infinite Kolam.

\subsection{Metric}

The polynomial in link components is given by the non-diagonalized inner product of the upper and lower half diagrams acted by a horizontal set of operators in the middle. The number of one-line drawings is now found as a coefficient of the leading term in the polynomial. It is stressed here that by up-down symmetry of the family of the Kolam diagrams, it is sufficient to classify the upper half and the action of the operators on it. The matrix representation of the metric of the state space will give the answer. Here, the metric is given by the inner products of states. Since there are two state spaces for the upper and lower half diagrams, the metric takes the matrix representation. Once we obtain it, it is easy to pick up the combinations of the upper and lower states which contribute to the infinite Kolam counting. In the following, we show this explicitly in the (1-3-5-3-1) diamond Kolam case. 

For the (1-3-5-3-1) diamond Kolam, the total number of Kolam patterns are given by $2^{16} = 65,536$. On the other hand, the total number of diagrams in $\bra{{\rm Upper}_5}$ is $2^6 = 64$, while the dimension of the state space is $945$. If one excludes the degree of freedom of the top black site, since \diaC at this site is irrelevant to the infinite Kolam counting, then the number of diagrams in $\bra{{\rm Upper}_5}$ is $2^5 = 32$. Similarly, if one excludes the two irrelevant caps in the left and right ends of the diagram, the dimension of the state space becomes $15$.

The explicit form of the $\bra{{\rm Upper}_5}$ is given by
\bea
  \bra{{\rm Upper}_5}
  &=&
  \bra{TOP_5} \diaB_{5} \diaW_{4} \diaW_{6} \diaB_{3} \diaB_{5} \diaB_{7}
  \nn
  &=& \bra{\cap_{1} \cap_{3} \cap_{5} \cap_{7} \cap_{9} \diaB_{5} \diaW_{4} \diaW_{6} \diaB_{3} \diaB_{5} \diaB_{7} } , 
\eea
where $\cap_i$ or {\small $\diaCap_{\, i}$} denotes {\small $\diaCap_{\hspace{-12pt}i ~~~i+1}$}. Any operator of the form $O_{i}$ denotes $~O_{\hspace{-10pt} i ~~ i+1}$ in what follows. In order to reduce the task for our purpose, we eliminate the $\diaB_5$ at the top black site and $\cap_1 \cap_9$.
\bea
  \bra{{\rm Upper}_5}_{a=0}
  &=&
  \bra{\cap_{3} \cap_{5} \cap_{7} \diaW_{4} \diaW_{6} \diaB_{3} \diaB_{5} \diaB_{7} }_{a=0} 
  \nn
  &=& 
  \bra{\cap_{5} \left( \cap_{3} \diaW_{4} \diaB_{3} \right) \left( \cap_{7} \diaW_{6} \diaB_{7} \right) \diaB_{5} }_{a=0}
  , 
\eea
where
\bea
 \cap_{3} \diaW_{4} \diaB_{3}
  &\!\!=& \!\!\! 
  \diaCap_{3} \Cross_{3} + \diaCap_{3} \diaC_{3} + \diaCap_{3} \Cross_{4} \Cross_{3} + \diaCap_{3} \Cross_{4} \diaC_{3}
  \;=\;
  (2+a)\; \diaCap_{3} + \diaD^{\hspace{-12pt}5}_{\;3} \diaCap_{4} , 
\nn
\cap_{7} \diaW_{6} \diaB_{7}  
 &\!\!=& (2+a)\; \diaCap_{7} + \diaD^{\hspace{-12pt}6}_{\;8} \diaCap_{6} .
\eea
Accordingly,
\bea&&\hspace{-25pt}
  \bra{{\rm Upper}_5}_{a=0}
\nn
  &=&
  \bra{\cap_{5} \left( 2 \cap_{3} + \diaD^{\hspace{-12pt}5}_{\;3} \cap_{4} \right) \left( 2 \cap_{7} + \diaD^{\hspace{-12pt}6}_{\;8} \cap_{6} \right) 
   \left( \Cross_{5} + \diaC_{5} \right) }_{a=0}
\nn
 &=&
 \bra{ \left( 4 \cap_{3} \cap_{5} \cap_{7} + 2\; \diaCap_{\hspace{-12pt} 3~~~6} \cap_{4} \cap_{7} + 2 \cap_{3} \diaCap_{\hspace{-12pt} 5~~~8} \cap_{6} + \diaCap_{\hspace{-12pt} 3~~~8} \cap_{4} \cap_{6} \right) \left( \Cross_{5} + \diaC_{5} \right)
 }_{a=0}
\nn
 &=&
 \bra{ 8 \cap_{3} \cap_{5} \cap_{7}
  + 2\; \diaCap_{\hspace{-12pt} 3~~~5} \diaCap_{\hspace{-12pt} 4~~~6} \cap_{7} 
  + 2 \cap_{3} \diaCap_{\hspace{-12pt} 5~~~7}\; \diaCap_{\hspace{-12pt} 6~~~8} 
  + \diaCap_{\hspace{-12pt} 3~~~8}\; \diaCap_{\hspace{-12pt} 4~~~6}\; \diaCap_{\hspace{-12pt} 5~~~7} 
  + \diaCap_{\hspace{-12pt} 3~~~8}\; \diaCap_{\hspace{-12pt} 4~~~7}\; \cap_{5}
 }.\nn[-5pt]
\eea
Hence, only 14 diagrams out of 32 and 5 distinct states out of 15 are relevant for the infinite Kolam counting. The diagrams are shown in Fig.5.
\begin{figure}[h]
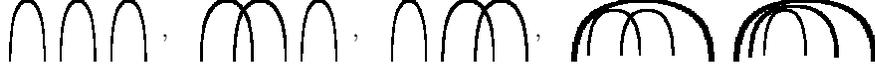

\hspace{18pt}
\scalebox{1.0}[2.0]{$\diaCap ~~\diaCap ~~\diaCap$}~ , ~~
\scalebox{1.5}[2.0]{$\diaCap \!\!\! \diaCap$}~ \scalebox{1.0}[2.0]{\diaCap} , ~~
\scalebox{1.0}[2.0]{\diaCap} \scalebox{1.5}[2.0]{$\diaCap \!\!\! \diaCap$} , ~~
\scalebox{4.0}[2.0]{\diaCap} \hspace{-65pt}\vspace{2pt} \scalebox{1.5}[1.5]{$\diaCap \!\!\! \diaCap$}~~~ , ~
\scalebox{4.0}[2.0]{\diaCap} \hspace{-65pt}\vspace{2pt} \scalebox{2.5}[1.7]{\diaCap} \hspace{-41pt}\vspace{2pt} \scalebox{1.2}[1.5]{\diaCap}
\caption{The five distinct states of the upper half diagram.}
\end{figure}

Next, we investigate the action of the white circle operators in the middle of the diagram onto the upper half diagram. It turns out that the action generates one more state into the expression.
\bea&&\hspace{-35pt}
  \bra{{\rm Upper}_5} \diaW_2 \diaW_4 \diaW_6 \diaW_8 \Bigr|_{a=0}
  \nn
  &=&
  \bra{{\rm Upper}_5}_{a=0} \diaW_4 \diaW_6 \Bigr|_{a=0}
  \nn
  &=&
   \bra{ 8 \cap_{3} \cap_{5} \cap_{7}
  + 2\; \diaCap_{\hspace{-12pt} 3~~~5} \diaCap_{\hspace{-12pt} 4~~~6} \cap_{7} 
  + 2 \cap_{3} \diaCap_{\hspace{-12pt} 5~~~7}\; \diaCap_{\hspace{-12pt} 6~~~8} 
  + \diaCap_{\hspace{-12pt} 3~~~8}\; \diaCap_{\hspace{-12pt} 4~~~6}\; \diaCap_{\hspace{-12pt} 5~~~7} 
  + \diaCap_{\hspace{-12pt} 3~~~8}\; \diaCap_{\hspace{-12pt} 4~~~7}\; \cap_{5}
 }  \diaW_4 \diaW_6 \Bigr|_{a=0}
 \nn
 &=& 
 \langle\; 12 \cap_{3} \cap_{5} \cap_{7}
  + 12 \; \diaCap_{\hspace{-12pt} 3~~~5} \diaCap_{\hspace{-12pt} 4~~~6} \cap_{7} 
  + 12 \cap_{3} \diaCap_{\hspace{-12pt} 5~~~7}\; \diaCap_{\hspace{-12pt} 6~~~8} \nn&&\qquad
  + 4\; \diaCap_{\hspace{-12pt} 3~~~8}\; \diaCap_{\hspace{-12pt} 4~~~6}\; \diaCap_{\hspace{-12pt} 5~~~7} 
  + 4\; \diaCap_{\hspace{-12pt} 3~~~8}\; \diaCap_{\hspace{-12pt} 4~~~7}\; \cap_{5}
  + 12\; \diaCap_{\hspace{-12pt} 3~~~5}\; \diaCap_{\hspace{-12pt} 4~~~7}\; \diaCap_{\hspace{-12pt} 6~~~8} \Bigr| . 
\eea

Finally, we calculate the metric of these six distinct states. If we label the states as:
\bea
\bra{1} &=& \bra{\cap_{3} \cap_{5} \cap_{7}} , 
\nn
\bra{2} &=& \bra{\diaCap_{\hspace{-12pt} 3~~~5} \diaCap_{\hspace{-12pt} 4~~~6} \cap_{7}} , 
\nn
\bra{3} &=& \bra{\cap_{3} \diaCap_{\hspace{-12pt} 5~~~7}\; \diaCap_{\hspace{-12pt} 6~~~8}} , 
\nn
\bra{4} &=& \bra{\diaCap_{\hspace{-12pt} 3~~~8}\; \diaCap_{\hspace{-12pt} 4~~~6}\; \diaCap_{\hspace{-12pt} 5~~~7}} , 
\nn
\bra{5} &=& \bra{\diaCap_{\hspace{-12pt} 3~~~8}\; \diaCap_{\hspace{-12pt} 4~~~7}\; \cap_{5}} , 
\nn
\bra{6} &=& \bra{\diaCap_{\hspace{-12pt} 3~~~5}\; \diaCap_{\hspace{-12pt} 4~~~7}\; \diaCap_{\hspace{-12pt} 6~~~8}} , 
\eea
the metric $G^{(5)}_{ij}$ is given by the inner products:
\bea
  G^{(5)}_{ij} &=& \braket{i \,|\, j} 
  = a \times \left(
  \begin{array}{cccccc}
  a^2 & a & a & 1 & a & 1 \\
  a & a^2 & 1 & a & 1 & a \\
  a & 1 & a^2 & a & 1 & a \\
  1 & a & a & a^2 & a & 1 \\
  a & 1 & 1 & a & a^2 & a \\
  1 & a & a & 1 & a & a^2 
  \end{array}
  \right) , 
\nn
  {G^\prime}^{(5)}_{ij} \Bigr|_{a=0}
  &=& \left(
  \begin{array}{cccccc}
  0 & 0 & 0 & 1 & 0 & 1 \\
  0 & 0 & 1 & 0 & 1 & 0 \\
  0 & 1 & 0 & 0 & 1 & 0 \\
  1 & 0 & 0 & 0 & 0 & 1 \\
  0 & 1 & 1 & 0 & 0 & 0 \\
  1 & 0 & 0 & 1 & 0 & 0 
  \end{array}
  \right) , 
\eea
where
$\ket{j}$ is the transposition of $\bra{j}$. The prime on $G$ stands for the partial differentiation by $\pa/\pa a$. By construction, the metric is symmetric. Note that the diagonal elements of the metric $G$ are always $a^{n/2}$ where $n$ is the number of legs.

\section{The results}

In general, when one obtains the upper half diagram as
\bea
\bra{{\rm Upper}_{N}}_{a=0} = \sum_{i=1}^{m} x_i \bra{i} , 
\eea
and its transformed one as
\bea
\bra{{\rm Upper}_{N}} \diaW^{N-1} \bigr|_{a=0} = \sum_{i=1}^{n\geq m} y_i \bra{i} , 
\eea
the answer to Problem 2.1 is given by the coefficients $x_i$, $y_i$, and the metric:
\bea
  P^\prime [ {\rm Kolam} ]_{a=0} &=& \sum_{i=1}^{n\geq m} \sum_{j=1}^{m}  x_i y_i G^\prime_{ij} \bigr|_{a=0} = {\mathbf y}^T G^\prime {\mathbf x} |_{a=0} , 
\eea
where $\{\bra{i}\}$ are the set of independent states, the prime denotes the partial derivative in $a$. ${\mathbf x}$ and ${\mathbf y}$ are the vectors though $x_{i>m}$ are all empty.

For the (1-3-1) Kolam patterns, the number of states in the upper half diagram is trivially given by one for our problem. Therefore, the number of the infinite Kolam is one.

For the (1-3-5-3-1) Kolam patterns, which is already complicated for human power, ${\mathbf x}$ is $(8,2,2,1,1,0)$ and ${\mathbf y}$ is $(12,12,12,4,4,12)$. Substituting these and the metric (23) into Eq.(26), we obtain
\bea
  P^\prime [\mbox{(1-3-5-3-1)}]_{a=0} &=& {\mathbf y}^T G^\prime {\mathbf x} |_{a=0}
  \;=\; 240 . 
\eea

In the same way, for the (1-3-5-7-5-3-1) case, 
 the number of states in the upper half diagram is 2,368 for infinite Kolam counting, while there are only 43 distinct basis states. The answer to Problem 2.1 is found to be 11,661,312. This means the infinite Kolam is very rare since it is 0.01696944 \% of the total number of Kolam patterns.

\section{Conclusion and discussions}

We have explicitly shown how to describe and analyze the Kolam in knot theory. We focused on the infinite Kolam counting, the notion of state and state space were introduced. With them, it was revealed that the corresponding metric calculation gives the answer to Problem 2.1. Especially, in the first few examples of the diamond Kolam, we have obtained the answers to Problem 2.1. Namely, 
for the grid (1-3-5-3-1), 
the answer is 240. 
For the (1-3-5-7-5-3-1), 
it is 11,661,312. They match the results in YANAGISAWA(2006) and another one by Yanagisawa.
Unfortunately, we have not reached the general solution of Problem 2.1 for the diamond Kolam (1-3-...-(2N+1)-...-3-1).

During this calculation, we have observed that the computational effort is much reduced because, for example in the (1-3-5-3-1) case, the number of the possible basis states is 15, whereas we obtain only five for the upper half and six for the transformed one. The degree of freedom of $n\times n$ symmetric metric is $n(n+1)/2$. Since the result is given by the metric, the steps of calculation is given by its degree of freedom and is roughly reduced from $15\cdot 16/2 = 120$ to $6\cdot 7/2 = 21$. If we compare with the total number of diagrams $2^{16}$ in the (1-3-5-3-1), our calculation ended dramatically with only $21$ cases. Note that for the state calculation, we set $a=0$. This is equivalent to have the grassmannian $a$ in the polynomial, which certainly reduced the task as was mentioned earlier.
Hence, we conclude that the calculation can be much more reduced than to count all the cases, and conjecture that the problem is not NP complete. 

Problem 2.1 is not a search problem of a particular diagram, so it should be warned that Problem 2.1 is neither of class NP nor of any class of search problem. However, the counting problem implicitly contains search problems. For example, if we assign different contour lengths for \Smooth and \Cross, we can ask the paths shorter than a certain distance, or the shortest paths. Piling up such information guides us to the answer to Problem 2.1. Such a component of Problem 2.1 resembles the famous problem called the ``Travelling Salesman Problem (TSP)''. In our case, the cities are replaced by the middle point of the edges in the initial lattice and the distances are assigned between them appropriately. The original TSP is known to be NP-hard. On the other hand, we have shown that at each step of calculation we can judge which vertex be taken for the solution. So, it is inferred that the algorithm we developed and the formulas we can develop find the solution.
Our method gave a sketch of the proof for our conjecture 
in this sense.

Finally, it should be remarked that
we have not discussed the symmetries of the Kolam, so our answers do not reflect them. Also, we report that some general statements and theorems on the state space were found together in regards to symmetries, which do not fit to this paper. Their details are to be reported elsewhere.

\vspace*{10pt}
\noindent
{\bf Acknowledgments}

{The author is grateful to K. Yanagisawa for his giving me this problem.
The author would also like to thank S. Nagata and T. Robinson for their comments.}

\appendix
\vspace*{20pt}
\section*{Appendix A$\quad$ The modified Temperley-Lieb algebra}

The modified Temperley-Lieb algebra is given by the Temperley-Lieb algebra (TL) with the decomposed cap and cup operators and the symmetry group which is reduced from the braid group. They are expressed by the following algebraic relations with some commutators:
\bea
&&
\cap_{i,j} = \cap_{j,i},\quad \cup^{i,j} = \cup^{j,i}, \quad
\sigma_{i,j} = \sigma_{j,i} ,   \nn&&
 \cap_{i,j} \sigma_{i,j} = \cap_{i,j} ,\quad 
 \sigma_{i,j} \cup^{i,j} = \cup^{i,j} ,\quad 
 \sigma_{i,j}^{~~2} = {\mathbf 1} \nn&&
 \cap_{i,j} \sigma_{j,k} = \delta^k_{\;j} \cap_{i,k}, \quad
 \sigma_{i,j} \cup^{j,k} = \cup^{i,k} \delta^j_{\;i}  \nn&&
 \cap_{ij}\, \delta^{j}_{\;k} = \cap_{ik},\quad \delta^{i}_{\;j}\, \cup^{jk} = \cup^{ik},   \nn&&
 \delta^i_{\;j}\sigma_{jk} = \delta^k_{\;j} \delta^i_{\;k} = \sigma_{ik} \delta^i_{\;j} , \nn&&
\cap_{i,j} \cup^{j,k} = \delta^{k}_{\;i} ,  \nn&&
\cap_{i,j} \cup^{i,j} = a \in \R  \nn&&
[ \cap_{i,j}, \cap_{k,l} ] = 0,\quad [ \cup^{i,j}, \cup^{k,l} ] = 0,\quad [ \cap_{i,j}, \cup^{k,l} ] = 0, \nn&&
[ \cup^{i,j}, \sigma_{k,l} ] = 0,\quad [ \cap_{i,j}, \sigma_{k,l} ] = 0, \quad
[ \sigma_{i,j}, \sigma_{k,l} ] = 0, \quad
for~~ \{i,j\}\wedge\{k,l\} = \emptyset. 
\eea
where $\sigma_{i,j}$ is an element of the symmetry (permutation) group which replaces the $i$-th entry with the $j$-th entry. The contraction of the suffixes is performed between the upper suffix on the right hand side and the lower suffix on its left.
Here, $a$ should not necessarily be a real number, but we write it for simplicity. 
The definition of $U_{i,j}$ is given by
\bea
  U_{i,j} = \cup^{i,j} \cap_{i,j} ,
\eea
so its algebra is also contained in the above. Note that $U_i \equiv U_{i,i+1}$ form the genuine Temperley-Lieb algebra.
With the above symbolic expressions of the algebra, the Reidemeister moves can be written down as below.\\

\underline{\bf Reidemeister moves}
\begin{enumerate}
 \def\theenumi{\roman{enumi}}
 \def\labelenumi{(\theenumi)}
 \item $\cap_{23} \sigma_1 \cup^{23} = \cap_{23} \cup^{13} \delta_{21} = \delta_{12} \delta_{21} =\delta_{11} =1
                         = 1_{11} (O_{23})$
\item 
$\cap_{23} \sigma_{1} \sigma_{3} \cup^{23} = \cup^{14} \cap_{14} = U_{14} (O_{23})$
\item not applicable to our problem.\\
\end{enumerate}

Finally, we write the black and white operators in terms of the mathematical symbols for convenience.
\bea
  \diaB &\equiv& \sigma_{i} + U_{i,i+1} \;,\;\; 
  \diaW \,\equiv\, {\mathbf 1} + \sigma_{i} \;,
\eea
where
$\sigma_{i} \equiv \sigma_{i,i+1}$. It is known that any order of the set $\{i\}$ can be represented by a product of $\sigma_{i=1,..,2n-1}$. Other basic facts of $\sigma_i$ are 
\bea&&
  \sigma_i^2 = 1 ,\; \sigma_i^{-1} = \sigma_i , 
\nn&&
  \sigma_{i+1} \sigma_{i} \sigma_{i+1} = \sigma_{i} \sigma_{i+1} \sigma_{i} , 
\nn&&
  [\sigma_i,\sigma_j] = 0 ~~~~for~~ |i-j|>1 .
\eea


\end{document}